# Griffin Plots of vortex-induced vibrations: revealing self-similarity for quick estimation from transient displacement responses

Guangzhong Gao[a*], Suhan Li[a], Jianming Hao[a], Bo Fu[b], Shucheng Yang[a], Ledong Zhu[c,d,e]

[a] Department of Bridge Engineering, Highway College, Chang'an University, 710064 Xi'an, China
[b] School of Civil Engineering, Chang'an University, Xi'an, 710064, China.
[c] Department of Bridge Engineering, Tongji University, 200092 Shanghai, China
[d] State Key Laboratory of Disaster Reduction in Civil Engineering, Tongji University, Shanghai,200092, China
[e] Key Laboratory of Transport Industry of Bridge Wind Resistance Technology, Tongji University, Shanghai, 200092, China

**Abstract:** Griffin plot relates the peak amplitudes of vortex-induced vibration to structrual mass-damping parameter, known as the Scruton number. Griffin plot serves as a fundamental tool in many engineering fields. This study confirms a general self-similarity in Griffin plots, where plots derived from transient responses at any Scruton number converge to a single, consisten curve. This self-similarity arises from weak aeroelastic nonlinearity in vortex-induced vibration, manifasting as amplitude-dependent aerodynamic damping. Based on this self-similarity property, we propose a numerical method to estimate Griffin plots from transient displacement responses at any Scruton number. The resulting plots align closely with experimental data for both cross-flow and torsional vortex-induced vibrations, highlighting robust self-similar behavior across different Scruton numbers. Furthermore, we observe a consistent trend in Griffin plots for a rectangular cylinder, closed-box, and double-girder bridge deck: the reciprocal of peak amplitudes shows an approximately linear relationship with the Scruton number, especially in torsional vortex-induced vibration. To generate this linearity, we develop a simple empirical model of vortex-induced forces. This model accurately reproduces the Griffin plot for a rectangular cylinder using aeroelastic parameters derived from a single Scruton number, significantly reducing the need for extensive experimental measurements.

## 1. Introduction

Vortex-induced vibration (VIV) is a fundamental challenge in engineering, impacting a wide range of structures and raising significant safety concerns. It affects the dynamics of riser tubes in offshore systems, flexible bridge components such as decks, hangers, stay cables, and arches, and civil engineering structures like chimneys, pillars, and lampposts, among other applications. The frequent occurrence of VIV and its engineering significance have made it a central topic in wind engineering community, attracting extensive investigation over recent decades (Li et al., 2024). Despite the existing studies, several fundamental questions remain unresolved, as reviewed by Williamson and Govardhan (2004, 2008) and Francisco (2025). Among these, a key issue is the relationship between maximum VIV amplitude and the structural mass and damping properties.

For a given shape profile, VIV peak amplitude is dependent on mass-damping parameters. Researchers have proposed various dimensionless parameters to characterize this dependency. Scruton (1963, 1965) introduced a dimensionless parameter, later named the *Scruton number* by Zdravkovich (1982), defined as $Sc = \pi m^* \xi_s / 2$, where $m^* = 4m/(\pi \rho D^2)$ is the mass ratio, $\rho$ is fluid density, $D$ is the cross-flow dimension, $m$ is the effective structural mass per unit length, $\xi_s$ is the structural damping ratio. This parameter effectively collapses peak amplitude data for elastically supported cylinders across different mass-damping conditions. Vickery and Watkins (1964) employed a similar dimensionless mass-damping parameter to plot peak VIV amplitudes for flexible cantilevers, referring to it as the *Stability parameter* $K_s = \pi^2 \left( m^* \xi_s \right)$. Skop and Griffin (1973) developed another comparable parameter, later termed the *Skop-Griffin parameter* by Skop (1974) to predict VIV amplitudes from compiled experimental data; the parameter is defined as $S_G = 2\pi^3 St^2 \left( m^* \xi_s \right)$, where *St* is the Strouhal number. Vandiver (2002) adapted the *Skop-Griffin parameter* for sheared flows by introducing a length ratio between the power-in region and the total length. Hansen (2013) proposed a general mass-damping parameter for non-cirucular sections, adjusting the Scruton number to be $Sc = 4\pi m / \left( \rho BD \right)$, where *B* is the cross-flow width, in order to account for the variation of relative flow velocities due to vortex shedding.

Historically, a long-standing debate has centered on whether structural mass and damping can be unified into a single parameter to collapse peak-amplitude data (Sarpkaya, 1979). Subsequent research, including Zdravkovich (1990), Williamson and Govardhan (2008), and Marra et al. (2015), suggests that for structures with large mass ratios, particularly in air flow ($m^* > 100$), peak VIV amplitudes depend solely on Scruton number. However, this conclusion primarily applies to cross-flow VIV, which has been the focus of most studies. Torsional VIV, a common phenomenon in civil engineering structures, remain underexplored, with limited consensus on its parametric dependencies.

In the literature, peak VIV amplitudes are commonly plotted against the Scruton number to compile experimental data. This approach is initially termed the *Skop-Griffin plot* and later simplified to the *Griffin plot* by Khalak and Williamson (1999). Subsequently, Govardhan and Williamson (2006) developed a *modified Griffin plot* that incorporates Reynolds number effects, significantly reducing the scatter in peak-amplitude data observed in the original Griffin plot of circular cylinders. The Griffin plot serves as an important engineering tool for assessing VIV responses in practical applications. Structural damping ratios and effective oscillating masses differ across vibration modes (Kim et al., 2016; Hwang et al., 2020), and even within a specific mode, damping can vary due to factors such as seasonal temperature fluctuations (Tan et al., 2024), material aging (Qu et al., 2024), or unexpected large-amplitude events (Ge et al., 2022). Consequently, evaluating peak VIV amplitudes for each mode requires considering various mass-damping combinations. The Griffin plot also plays a crucial role in designing external dampers for prototype bridges, enabling engineers to assess mitigation effects across different structural damping ratios. In wind tunnel experiments, Griffin plots are frequently employed to estimate VIV peak responses at the design damping level based on experimental damping, as precisely adjusting structural damping to the design level in section model tests is often impractical (Hwang et al., 2020).

Accurate measurement of the Griffin plot is a primary focus in experimental studies (Wu and Laima, 2021). Traditionally, Griffin plots are derived from elastically supported section model tests, which involve measuring peak VIV amplitudes across a large range of Scruton numbers and interpolating amplitudes for untested values using existing data points (Hwang et al., 2020). This method is time-intensive, particularly when incorporating variations in wind angle of attack (typically ranging from -5° to 5°) and the implementation of VIV mitigation countermeasures. An alternative, indirect approach to generating Griffin plots involves empirical models. Numerous empirical VIV models exist, broadly grouped into two categories: single-degree-of-freedom models and wake oscillator models. Single-degree-of-freedom models can be subdivided into: nonlinear damping models (Ehsan and Scanlan, 1990; Larsen, 1995; Zhu et al., 2017; Marra et al., 2015; Gao et al., 2021; Qie and Zhang, 2024), and force coefficient data models (Sarpkaya, 1978; Iwan and Botelho, 1985; Zasso et al., 2008). Wake oscillator models, on the other hand, introduce an independent equation to capture nonlinear wake dynamics (Hartlen and Currie, 1970; Skop and Griffin, 1973; Diana et al., 2006; Farshidianfar and Zanganeh, 2010; Chen et al., 2022; Hui et al., 2024). Several of these models have demonstrated the ability to reproduce Griffin plots by estimating parameters across various Scruton numbers (Larsen, 1995; Marra et al., 2015; Skop and Griffin, 1973; Farshidianfar and Zanganeh, 2010; Chen et al., 2022). Notably, Marra et al. (2011) assessed the famous Ehsan and Scanlan's model using experimental data from a rectangular cylinder and found that it failed to predict VIV amplitudes when Scruton numbers deviate from the value used for parameter identification. To address this limitation, Marra et al. (2015) improved the model by incorporating Scruton number dependency into the aeroelastic parameters. It requires data from at least three distinct Scruton numbers to estimate parameters for models such as Larsen's (1995) and Marra's refined version (2015). The ability to reproduce Griffin plots is a fundamental requirement for any empirical model, and an effective model should achieve this with minimal experimental measurement. It is also noteworthy that most existing models were developed primarily for cross-flow VIV, leaving their applicability to torsional VIV largely unverified.

In this study, we reveal an inherent self-similarity in Griffin plots for bluff bodies. The Griffin plot emerges from aeroelastic nonlinearity during VIV instability, and bluff body aeroelasticity typically exhibits weak nonlinearity, characterized by quasi-harmonic displacement responses and amplitude-dependent aerodynamic damping (Gao et al., 2021). For a given section profile, it is proved in this study that the Griffin plot can be derived from the amplitude-dependent aerodynamic damping estimated from transient displacement responses. Griffin plots constructed from various Scruton numbers collapse to experimental results obtained through conventional methods, thus revealing self-similar behavior. This self-similarity is validated using VIV experimental data from typical bluff sections. Additionally, we investigate the applicability of self-similarity and the Scruton number as a unified dimensionless mass-damping parameter for torsional VIV. Building on these findings, we propose a simple empirical model that requires experimental data from only one Scruton number, offering a practical solution for engineering applications. This study focuses on VIV in air flows and bluff bodies with sharp corners, where the flow field is insensitive to Reynolds number effects, and thus resulting Griffin plot is also insensitive to Reynolds number efffects.

## 2. Self-similarity in Griffin plots for vortex-induced vibrations

Griffin plots represent the relationship between the stable amplitude of vortex-induced vibrations (VIV) and the Scruton number, a dimensionless parameter that combines structural mass and damping. At the stable amplitude stage of VIV, aerodynamic damping balances structural damping, resulting in zero total damping. Thus, the Griffin plot also reflects how aerodynamic damping relates to the stable VIV amplitude.

In contrast, the amplitude-dependent behavior of aerodynamic damping can be derived from the transient displacement envelopes during the decay-to-resonace (DTR) and grow-to-resonance (GTR) processes.

Theoretical analysis confirms that, to a first-order approximation, the aerodynamic damping of a bluff body depends only on the oscillation amplitude (Gao et al., 2021; Zhang et al., 2024). In this context, the total damping ratio acts as the small paramerer. Other structural factors, including the Scrtuon number, have a higher-order, negligible effect on aerodynamic damping.

This observation suggests self-similarity in Griffin plots. Specifically, the Griffin plot for a bluff body can be approximated using the transient displacement envelopes at peak-amplitude states. When estimated across different Scruton numbers, these plots should collapse onto a single universal curve, consistent with the first-order approximation. Therefore, the Griffin plot can be estimated from transient displacement envelopes during DTR or GTR processes, regardless of the specific Scruton number, as long as it effectively describes how structural mass and damping influence peak VIV amplitudes.

## 3. Estimating Griffin plot using transient displacement envelopes

### 3.1. Numerical estimation procedure

The Griffin plot exhibits self-similarity, which enables its estimation from transient displacement responses. The following steps outline this estimation process in detail:

(1) Section model test setup and structural parameter identification

Configure the spring-suspended section model system to represent a representative mass-damping condition. In still air, apply an artificial perturbation to initiate a free decay oscillation. Record the resulting displacement signal and analyze it to determine the structural damping ratio $\xi_s$ and the natural oscillating frequency $f_0$ of the oscillatory system.

(2) Measurement of VIV lock-in region and transient responses

Gradually increase the oncoming wind velocity while monitoring the steady-state amplitudes of VIV within the lock-in region. Identify the wind speed at which the VIV amplitude reaches its maximum, whether for heaving or torsional modes. At this critical wind speed, artificially excite the section model to induce either a GTR response or a DTR response, where the transient amplitude evolves toward the steady-state value. The heaving displacement during these GTR or DTR responses can be expressed as:

$$y(t_i) = A(t_i)\cos(\omega t_i + \theta_0) \quad (1)$$

where $A(t_i)$ represents the time-varying amplitude envelope, $\omega$ is the oscillation frequency, and $\theta_0$ is the phase angle.

(3) Calculation of transient velocity responses

The velocity responses are derived from the recorded displacement signal using a high-order finite difference. The velocity $\dot{y}$ at time $t_i$ is computed as:

$$\dot{y}(t_i) = \begin{cases} \dfrac{-y(t_{i+2}) + 8y(t_{i+1}) - 8y(t_{i-1}) + y(t_{i-2})}{12\Delta t}, & 2 < i < N \\ \dfrac{-y(t_{i+2}) + 4y(t_{i+1}) - 3y(t_i)}{2\Delta t}, & 1 \leq i \leq 2 \\ \dfrac{3y(t_i) - 4y(t_{i-1}) + y(t_{i-2})}{2\Delta t}, & N-2 \leq i \leq N \end{cases} \quad (2)$$

where $N$ is the total number of sampled points, $\Delta t$ is the sampling time interval.

(4) Extraction of transient amplitude (displacement envelope)

The transient amplitude $A(t_i)$, also refer to as the displacement envelope, is calcualted from the displacement and velocity time histories using

$$A(t_i) = \sqrt{y(t_i)^2 + [\dot{y}(t_i)/\omega]^2} \quad (3)$$

where $\omega = 2\pi f$ is the circular frequency during VIV.

(5) Smoothing the transient amplitude

To remove the high frequency noise from the transient amplitude $A(t_i)$, the calculated data is fitted with respect to time $t$ using a trial function. A second-order exponential decay function is recommended

$$\bar{A}(t) = ae^{bt} + ce^{dt} \quad (4)$$

where $a, b, c, d$ are empirical constants, which are determined via least-square fitting to best match the calculated data $A(t_i)$.

(6) Calculation of total damping ratio

According to Gao et al. (2021), the total damping ratio $\xi_{\text{total}}(t_i)$ is calculated as

$$\xi_{\text{total}}(t_i) = -\frac{1}{\omega}\frac{d}{dt}\left(\ln \bar{A}\right)\bigg|_{t=t_i} = -\frac{1}{\omega}\frac{\dot{\bar{A}}(t_i)}{\bar{A}(t_i)} \tag{5}$$

where $\dot{\bar{A}}(t_i)$ is the time derivative of the fitted trial function from Step (5), which can be analytically expressed using the fitted constants $a, b, c, d$.

(7) Extraction of aerodynamic damping

According to Reference Gao et al. (2021), the amplitude-dependent aerodynamic damping $\xi_{\text{aero}}$ is obtained by subtracting the structural damping $\xi_s$ from the total damping:

$$\xi_{\text{aero}}(t_i) = -\frac{1}{\omega}\left[\frac{\dot{\bar{A}}(t_i)}{\bar{A}(t_i)} + \xi_s \omega_0\right] \tag{6}$$

Since $\xi_{\text{total}}(t_i)$ it computed at specific times corresponding to transient amplitude $A(t_i)$, this yields $\xi_{\text{aero}}$ as a function of transient amplitude $A$.

(8) Construction of the Griffin plot

The relationship between aerodynamic damping and amplitude is visualized by plotting $\xi_{\text{aero}}(t_i)$ versus transient amplitude $\bar{A}(t_i)$. To transform it into the Griffin plot, we further treat the transient amplitude as the stable amplitude, interpret the absolute value of $\xi_{\text{aero}}(\bar{A})$ as the structural damping ratio $\xi_s$ in steady-amplitude stage, and then multiply $\xi_s$ by the dimensionless mass to compute the Scruton number $Sc$.

For heaving VIV, the Scruton number is defined as

$$Sc = \frac{4\pi m \xi_s}{\rho BD} \tag{7}$$

For torsional VIV, it is

$$Sc = \frac{4\pi J_m \xi_s}{\rho B^3 D} \tag{8}$$

where $m$ and $J_m$ represent the effective mass and mass moment of inertia per unit length of the section model oscillatory system, respectively. $B$, $D$ are the width and depth of the cross section. $\rho$ is air density.

## 3.2. Griffin plots for heaving and torsional VIV of typical bluff sections

In this section, we validate the self-similarity of Griffin plots and the estimation method proposed in Section 3.1 using experimental data for typical bluff sections. The applicability of the method is assessed for both heaving and torsional VIV.

### 3.2.1 Heaving VIV of a rectangular section with side ratio 4:1

A rectangular section with a width-to-depth ratio of $B/D=4$ has been identified as prone to heaving VIV. Marra et al.(2015) measured the heaving VIV responses of this section across a wide range of Scruton number, using a magnetic damping device to adjust structural damping. Their experimental results are plotted in Fig.1.

To further evaluate the Griffin plots using our proposed method, we select three representative

Scruton numbers from the dataset, which are corresponding to the lowest tested *Sc* as 1.9, the medium *Sc* as 21.7, and the highest *Sc* as 78.1.

The DTR responses at these Scruton numbers are illustrated in Fig.2. The transient velocity was calculated using Eq.(2) and then the transient amplitudes were computed using Eq.(3). As shown in Fig.2, the computed amplitudes align well with the experimental displacement envelopes. The dynamic parameters associated with these Scruton numbers are listed in Table 3.

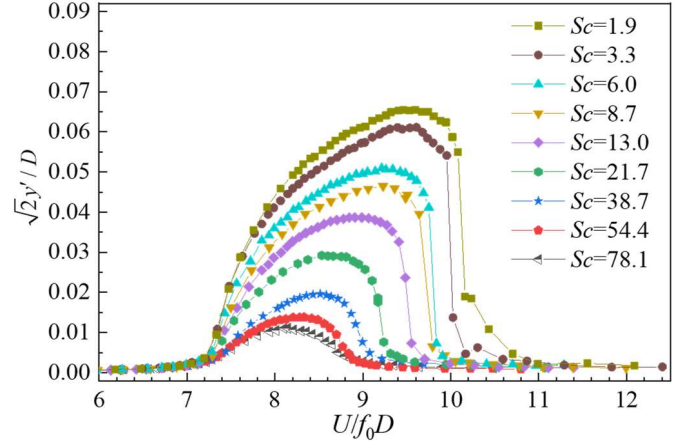

Fig. 1. Dimensionless VIV amplitude versus reduced velocity of a rectangular cylinder with side ratio $B/D$=4 at various Scruton numbers. $y'$ is the root mean square (RMS) of the steady-state displacement, and $\sqrt{2}y'/D$ is the dimensionless equivalent sinusoidal amplitudes (Marra et al.,2015).

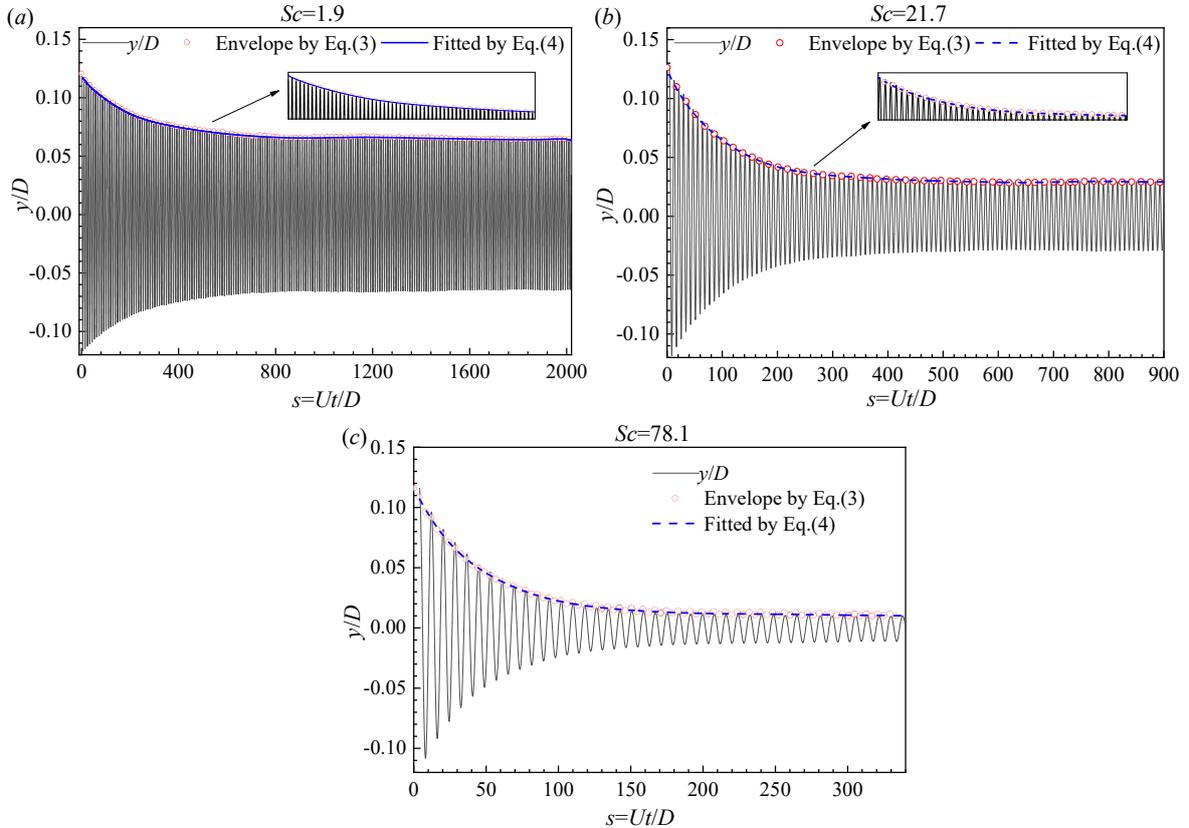

Fig. 2. Time histories of displacement during DTR responses for VIV of a rectangular cylinder with side ratio $B/D$=4 at three representative Scruton numbers. Data from Marra et al. (2015).

Table 1. Dynamic parameters for heaving VIV tests of a rectangular cylinder with side ratio $B/D$=4. Data from Marra et al. (2015).

| No. | $f_0$ (Hz) | $m$ (kg/m) | $\xi_s$ (%) | $Sc = \dfrac{4\pi m \xi_s}{\rho BD}$ | $\rho$ (kg/m$^3$) |
|---|---|---|---|---|---|
| 1# | 7.97 | 7.089 | 0.058 | 1.9 | 1.19 |
| 6# | 7.87 | 7.292 | 0.65 | 21.7 | 1.22 |
| 9# | 7.88 | 7.292 | 2.34 | 78.1 | 1.22 |

Following the estimation procedure outlined in the previous section, we investigate the relationship between aerodynamic damping and dimensionless amplitudes using three different DTR responses in Fig.2. Fig.3 plots aerodynamic damping as a function of dimensionless amplitude for three different Scruton numbers. Notably, the resulting curves exhibit good agreement, converging toward a consistent trend. This conherence supports the observation that aerodynamic damping depends primarily on transient amplitude and remain largely insensitive to the variation in structural mass-damping parameters, such as the Scruton number.

    Fig.4 presents the Griffin plots constructed from Fig.3, compared against experimental Griffin plot results derived from the peak amplitudes for each Scruton number shown in Fig.1. The constructed Griffin plots demonstrate remarkable consistency with the experimental data. However, the curves for different Scruton numbers span distinct amplitude ranges. This difference arises because lower Scruton numbers correspond to higher stable amplitude, which in turn result in smaller variation range of transient amplitude during the DTR process, as indicated in Fig.2. This limitation can be addressed by initiating the DTR process with a large initial amplitude and complementing it with GTR tests starting from rest during wind tunnel tests. Implementing these adjustments enables the estimated Griffin plot for a single Scruton number to cover a broader amplitude range. Additionally, Fig.4 includes negative Scruton numbers, which emerge from subtracting structural damping from the total damping. While these negative values lack physical meaning, thay are retained to illustrate the trend of the curves at very low Scruton number.

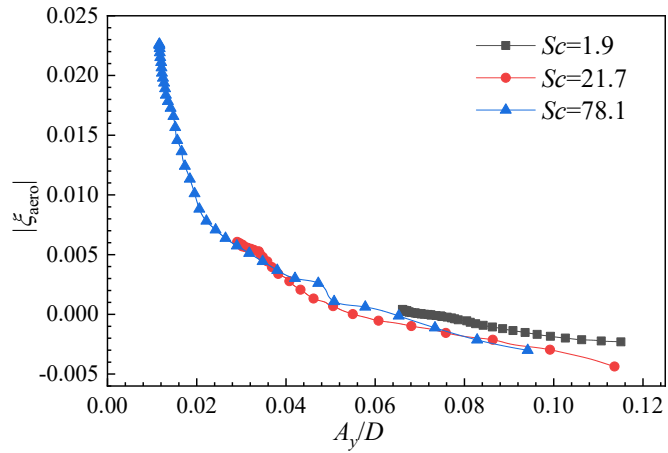

Fig. 3. Estimated aerodynamic damping ratio versus dimensionless amplitude for heaving VIV of a rectangular cylinder with side ratio $B/D$=4 at three Scruton numbers. $|\xi_{aero}|$ represents the absolute value of aerodynamic damping, $A_y / D$ is the dimensionless transient amplitude of heaving VIV.

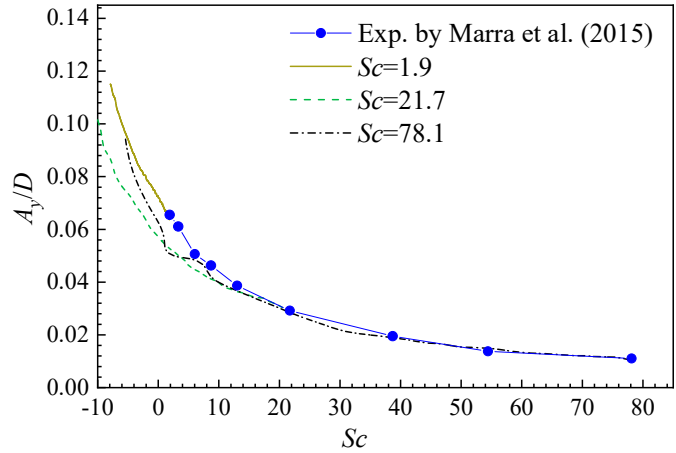

Fig. 4. Constracted Griffin plots for a rectangular cylinder with side ratio $B/D$=4 using second order exponential decay function as the trial function, compared with experimental data by Marra et al. (2015).The Solid, dashed, and dot-dash lines represent the Griffin plots derived from the corresponding aerodynamic damping curves at three different Scruton numbers in Fig.3.

### 3.2.2. Heaving and torsional VIV of a box-type bridge deck

The closed-box bridge deck, a common deck type, is susceptible to both heaving and torsional VIV. Spring-suspended section model tests were conducted on the closed-box bridge deck of the Humen Bridge (see Fig. 5) at a geometric scale of 1:60. The prototype bridge experienced sustained heaving VIV starting from the afternoon of May 5, 2020, as reported by Ge et al. (2022). With an initial wind angle of attack of +3°, heaving VIV was observed for reduced velocities ranging from 8 to 11 (see Fig.6a), while torsional VIV occurred for reduced velocities between 10 and 14 (see Fig.7a).

To examine the influence of structural mass-damping conditions, the structural damping ratios were adjusted during the section model tests to vary the Scruton number. The dynamic parameters for the heaving and torsional VIV experiments are presented in Table 2 and Table 3, respectively. Additionally, Table 2 incorporates variations in the dimensionless structural mass, enabling the exploration of Scruton numbers derived from different combinations of structural damping and mass parameters.

Using the proposed estimation procedure, the aerodynamic damping was determined as a function of dimensionless amplitude for various Scruton numbers. This analysis utilized grow-to-resonance (GTR) responses at peak VIV amplitudes. The results for heaving VIV are shown in Fig.6b, and for torsional VIV in Fig.7b. The aerodynamic damping curves for both heaving and torsional VIV exhibit strong consistency across different Scruton numbers, highlighting the existence of self-similarity in the amplitude-dependent aerodynamic damping.

Griffin plots for heaving and torsional VIV, constructed from transient displacement data, are displayed in Fig.6c and Fig.7c, respectively. These plots align closely with one another and show excellent agreement with experimental results derived from peak amplitudes in Fig.6a and Fig.7a. This consistency underscores the reliability of the estimation method and also highlights the self-similarity in Griffin plots.

Moreover, the results in Fig.6 demonstrate that the self-similarity in Griffin plots and the effectiveness of the proposed estimation method hold even when Scruton numbers are obtained from varying combinations of structural damping and dimensionless mass. As a result, this approach facilitates rapid estimation of the Griffin plot from responses under a single Scruton number.

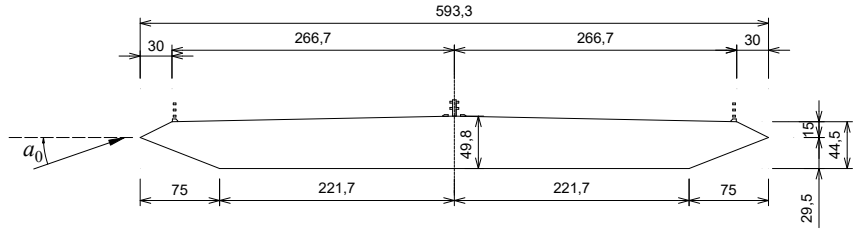

Fig. 5. Cross section of a closed-box bridge deck with an initial wind angle of attack $\alpha_0$. Dimensions are in millimeters.

Table 2. Dynamic parameters for heaving VIV tests on the closed-box bridge deck section of the Humen Bridge.

| No. | $f$ (Hz) | $m$ (kg/m) | $\xi_s$ (%) | $Sc = \dfrac{4\pi m \xi_s}{\rho BD}$ | $m^* = \dfrac{4m}{\pi \rho BD}$ |
|---|---|---|---|---|---|
| 1# | 4.248 | 4.892 | 0.305 | 5.2 | 172.4 |
| 2# | 4.346 | 4.892 | 0.442 | 7.5 | 172.4 |
| 3# | 4.492 | 4.892 | 0.624 | 10.6 | 172.4 |
| 4# | 4.492 | 4.892 | 0.859 | 14.6 | 172.4 |
| 5# | 5.078 | 4.892 | 2.113 | 36.0 | 172.4 |
| 6# | 4.199 | 5.859 | 0.874 | 17.8 | 206.4 |
| 7# | 4.102 | 5.859 | 0.437 | 8.9 | 206.4 |

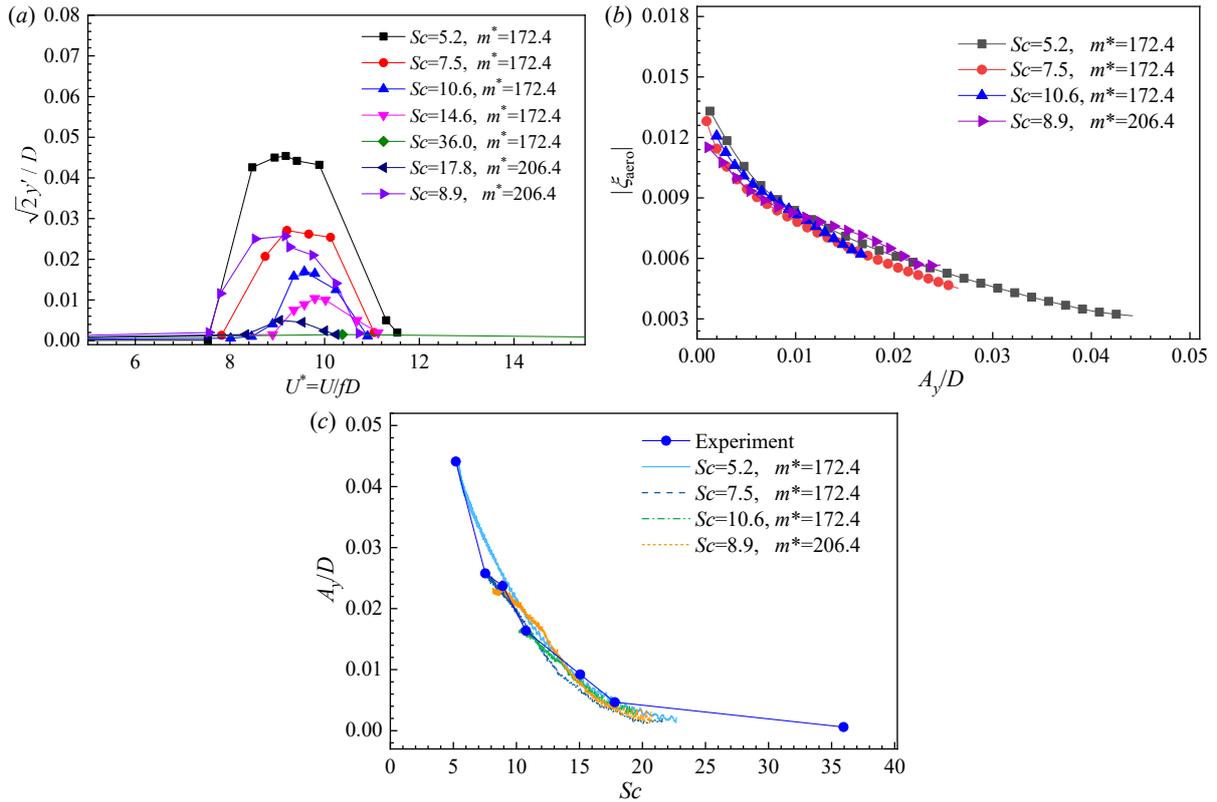

Fig. 6. Heaving VIV of the closed-box bridge deck section of the Humen Bridge. (a) Dimensionless VIV amplitude versus reduced wind velocity for various Scruton numbers and dimensionless mass, (b) Amplitude-dependent aerodynamic damping identified from GTR processes, (c) Constructed Griffin plots compared with experimental results. The solid, dashed, dot-dash and dash lines represent aerodynamic damping and Griffin plots derived from transient displacements at different Scruton numbers and dimensionless mass.

Table 3. Dynamic parameters for torsional VIV tests on the closed-box bridge deck section of the Humen Bridge

| No. | $f$ (Hz) | $J_m$ (kg·m$^2$/m) | $\xi_s$ (%) | $Sc = \dfrac{4\pi J_m \xi_s}{\rho B^3 D}$ | $m^* = \dfrac{J_m}{\pi \rho B^3 D}$ |
|---|---|---|---|---|---|
| 1# | 7.257 | 0.183 | 0.421 | 0.771 | 4.561 |
| 2# | 7.257 | 0.183 | 0.443 | 0.807 | 4.561 |
| 3# | 7.520 | 0.183 | 0.624 | 1.139 | 4.561 |

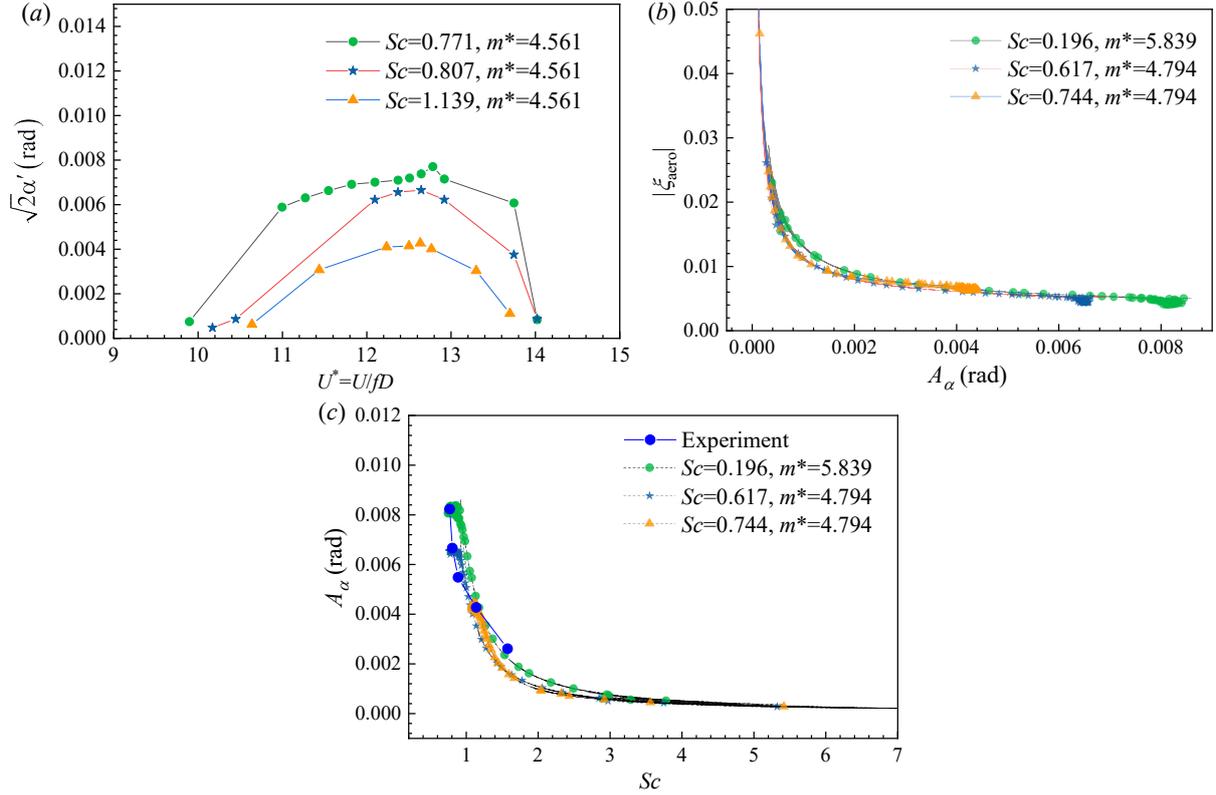

Fig.7. Torsional VIV response of the closed-box bridge deck section of the Humen Bridge. (a) TorsionalVIV amplitude versus reduced velocity for different Scruton numbers, (b) Amplitude-dependent aerodynamic damping identified from GTR processes, (c) Constructed Griffin plots compared with experimental results.

### 3.2.3. Torsional VIV of a double-girder bridge deck

The double-girder bridge deck is another common deck type susceptible to VIV. Gao et al. (2020, 2021) conducted spring-suspended section model tests on a representative double-girder deck section, as shown in Fig.8. The prototype for these experiments is the 500-meter PC Cable-Stayed North Branch Navigation Channel Bridge, part of the Jingzhou Yangtze River Bridge, with an experimental geometric scale of 1:25. At an initial wind angle of attack of 0°, torsional VIV was observed across reduced velocities ranging from 8 to 16.5 (see Fig.9a).

The dynamic parameters for these torsional VIV experiments are presented in Table 3. The Scruton number varies through adjustments in structural damping and mass moment of inertia, enabling an analysis of how different combinations of these parameters influence torsional VIV.

Fig.9b illustrates the aerodynamic damping as a function of dimensionless transient amplitude, estimated from displacement GTR responses using the numerical procedure described in Section 3.1. The aerodynamic damping curves demonstrate strong consistency across various Scruton numbers and dimensionless masses, confirming the self-similarity of torsional VIV. Corresponding Griffin plots, constructed from amplitude-dependent aerodynamic damping and validated against direct experimental results, are shown in Fig.9c. These plots exhibit excellent agreement, further supporting the reliability of the findings.

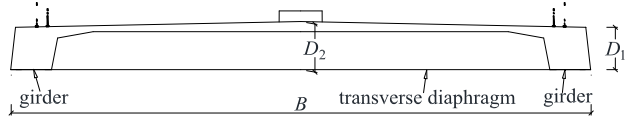

Fig.8. Cross section of a double-girder bridge deck with $B$=1.08m, $D_1$=0.08m, $D_2$=0.0908m. Transverse diaphragms are spaced at 0.32m intervals along the bridge's longitudinal axis (Gao et al., 2020).

Table 4. Dynamic parameters of torsional VIV tests on a typical double-girder bridge section.

| No. | $f$ (Hz) | $J_m$ (kg·m²/m) | $\xi_s$ (%) | $Sc = \dfrac{4\pi J_m \xi_s}{\rho B^3 D}$ | $m^* = \dfrac{J_m}{\pi \rho B^3 D}$ |
|---|---|---|---|---|---|
| 1# | 3.162 | 2.110 | 0.0960 | 0.182 | 4.794 |
| 2# | 3.174 | 2.115 | 0.326 | 0.617 | 4.794 |
| 3# | 3.174 | 2.115 | 0.393 | 0.744 | 4.794 |
| 4# | 2.930 | 2.576 | 0.0850 | 0.196 | 5.839 |

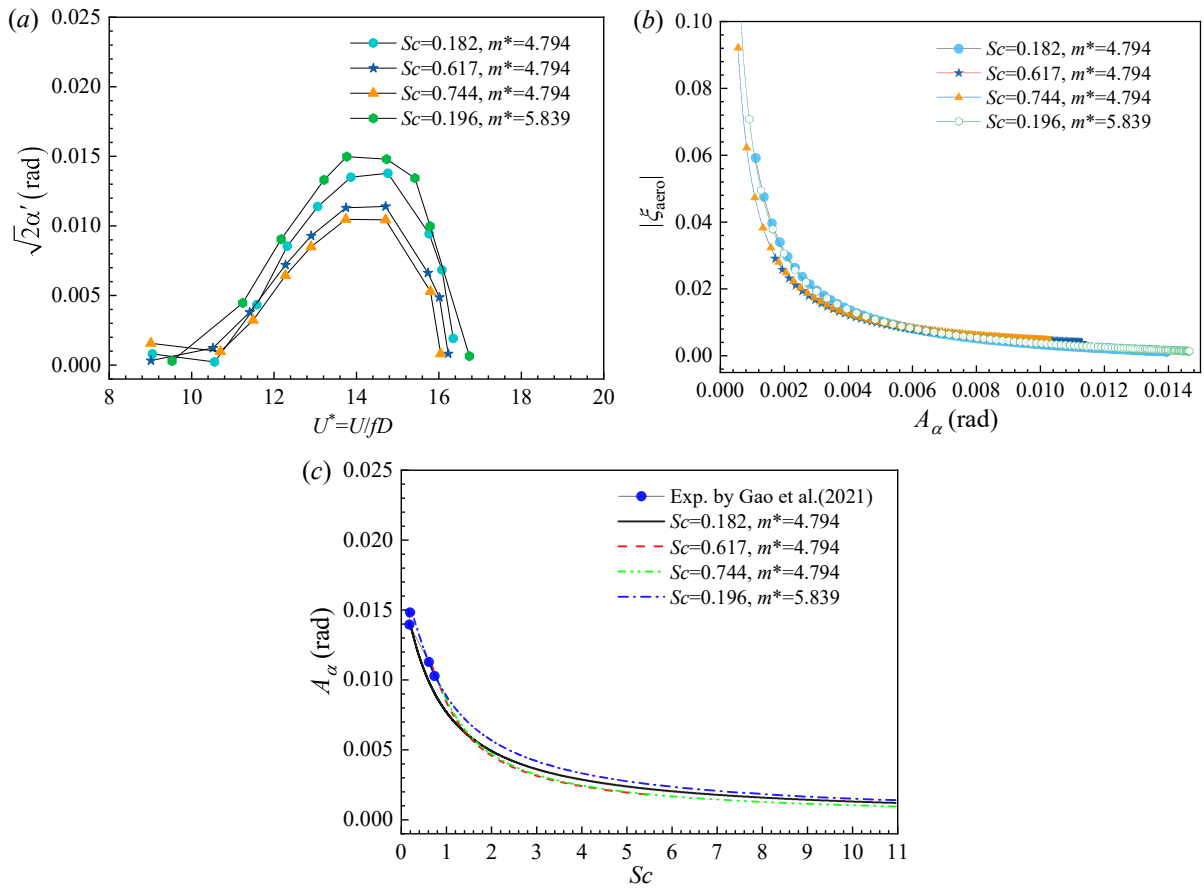

Fig. 9. Torsional VIV response of a typical double-girder bridge section. (a) Torsional amplitude versus as a function of reduced velocity for various Scruton numbers and dimesionless masses, (b) Amplitude-dependent aerodynamic damping identified from GTR processes, (c) Constructed Griffin plots compared with experimental results.

### 3.3. Characteristics of Griffin plots

Based on the estimated and experimental Griffin plots presented in Section 3.2, we further investigate the characteristics of Griffin plots to explore whether a general qualitative form exists. An intriguing observation for both heaving and torsional VIV is that the reciprocal of the stable amplitude exhibits an approximately linear relationship with the Scruton number across a wide range of amplitudes.

Fig.10 examines this linear relationship for the Griffin plots of heaving VIV (from Fig.4) and torsional VIV (from Fig.9c). For both the constructed Griffin plots and the experimental data, a clear linear relationship is evident between the reciprocal of the stable amplitude and the Scruton number. In Fig.10b, discrepancies among the constructed Griffin plot lines increase with the Scruton number, particularly at very small amplitudes where estimation errors are more pronounced. These errors, though exaggerated in the reciprocal plot, correspond to very small discrepancies in amplitude due to the nature of reciprocity. Notably, the lines in Fig.10b diverge when the amplitude falls below 0.286°.

Fig.11 further explores this linear relationship for the heaving VIV of the Humen Bridge (from Fig.6c). Both the constructed Griffin plots and experimental data points exhibit a clear linear trend for Scruton numbers below approximately 13. However, deviation from linearity occurs for Scruton numbers above 13, attributed to Coulomb friction effects from external dampers used in the wind tunnel experiments. Coulomb friction was employed to increase structural damping in the spring-suspended section model tests, and prior studies (Gao and Zhu, 2015) have shown that heaving-mode damping is susceptible to this effect, leading to an enhanced amplitude-dependent behavior at small amplitudes. In contrast, torsional-mode damping remains largely unaffected by friction. Despite this, the linear relationship remains a satisfactory approximation, as the influence of Coulomb friction is confined to very small amplitudes. As shown in Fig.11, deviation begins at a Scruton number of 13.2, corresponding to a stable amplitude of approximately $D/80$, which is a very small value for long-span bridges.

This linear relationship is further validated in Fig.12 using experimental results from the heaving and torsional VIV of the steel pylon of the 1915 Çanakkale Bridge, measured in aeroelastic model tests (Liao, 2021). The VIV of an aeroelastic model can be treated as an equivalent single mode vibration. A clear linear trend between the reciprocal of the stable amplitude and the structural damping ratio is observed for torsional VIV. However, the heaving VIV curve tilts upwards, again due to the Coulomb friction effect, consistent with previous findings.

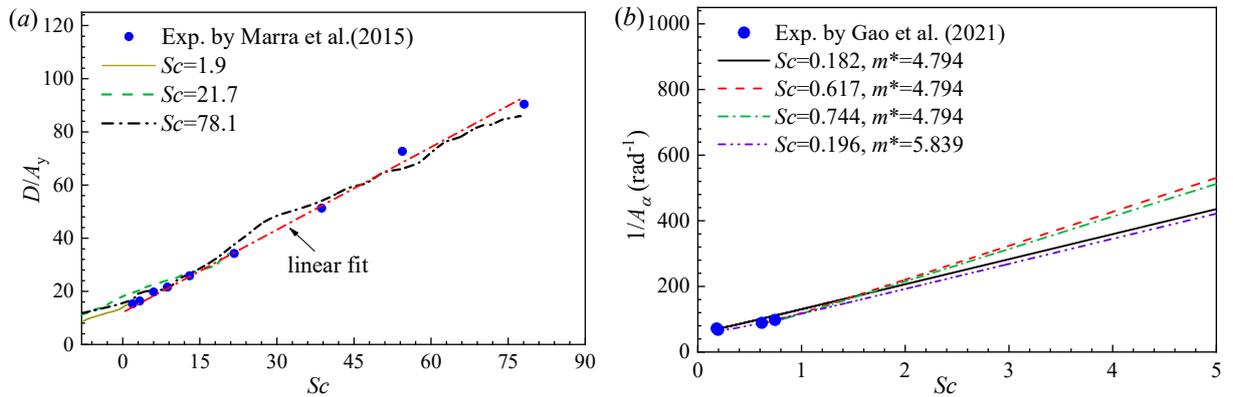

Fig. 10. Linear relationship between the Scruton number and the reciprocal of stable amplitudes. (a) Heaving VIV of a rectangular 4:1 cylinder, and this figure is transformed from Fig.4. (b) Torsional VIV of a double-girder section, and this figure is transformed from Fig.9c. Lines represent the constructed data of Griffin plots under the responding structural mass-damping conditions in the legends.

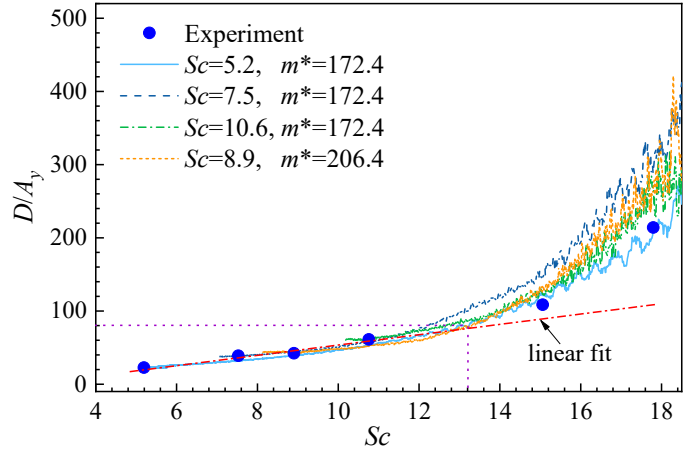

Fig. 11. The influence of Columb friction under small-amplitude regime on the linear relationship for the heaving VIV of a closed-box girder. This figure is transformed from Fig.6c.

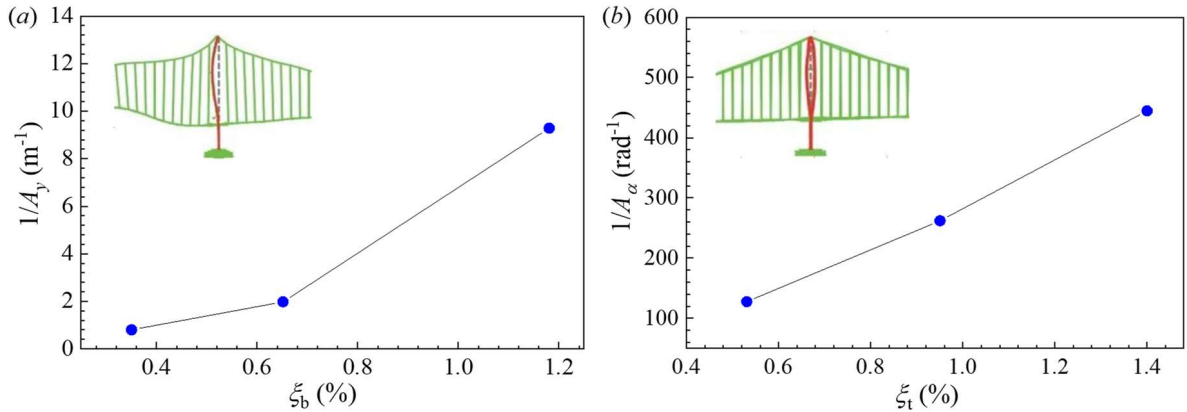

Fig. 12. The relationship between damping ratio and the reciprocal of stable amplitude for the VIV of 1915 Çanakkale Bridge pylon.(a) Bending VIV of bridge pylon along deck axis, (b) torsional VIV of bridge pylon. Data from Liao (2021).

### 3.4. Discussion on the estimation of Griffin plots

The self-similarity and estimation method for Griffin plots have been validated for both heaving and torsional VIV across typical bluff sections, as detailed in Section 3.2. This self-similarity carries significant physical meaning: the transient displacement envelope at any structural mass-damping condition contains enough information to reconstruct a general Griffin plot applicable to other mass-damping combinations.

The following aspects are worthy to be further discussed.

(1) Traditional approach vs. self-similarity

Traditionally, obtaining a Griffin plot requires extensive wind tunnel experiments across a broad range of structural mass-damping conditions. Stable amplitudes at untested Scruton numbers are then interpolated from existing data, necessitating small intervals of tested Scruton numbers to minimize interpolation errors. This process is both time-consuming and cumbersome. However, by using self-similarity, a Griffin plot can be efficiently reconstructed from a single set of transient displacement data at any Scruton number, enabling rapid estimation and extrapolation to other Scruton numbers.

(2) Extending self-similarity to other phenomena

Weak aerodynamic nonlinearity in VIV, galloping, and flutter has been widely observed across various bluff sections (Gao et al., 2021). This suggests that self-similarity is an inherent characteristic of bluff body aeroelasticity. Consequently, the proposed estimation method can be extended to predict nonlinear behaviors in soft flutter or galloping across different structural mass-damping parameters.

(3) Qualitative form and empirical modeling

Section 3.3 identifies a general qualitative form of Griffin plots for typical bluff sections: the reciprocal of stable VIV amplitudes increases approximately linearly with the Scruton number. Recalling from Gao and Zhu (2015) that Coulomb friction effect also results in a similar behavior with oscillation amplitude for the structural damping, it is speculated that separated flow fields exert a negative Coulomb friction effect on the oscillatory system of the bluff body. Building this linear relationship, an empirical model of vortex-induced forces or torques can be developed to preserve this linearity in the constructed Griffin plots, which will be explored in the next section. Notably, friction-type dampers in wind tunnel experiments can adversely affect the accuracy of Griffin plot estimations, particularly for heaving VIV, as shown in Fig.11. Magnetic (Marra et al., 2015), oil dampers (Gao and Zhu, 2015) or the damper based on op-amp circuit and controller (Govardhan and Williamson, 2009) are recommended for VIV experiments, as they provide consistent structural damping.

(4) Importance of selecting the trial function

Accurate calculation of amplitude-dependent aerodynamic damping is critical for estimating Griffin plots from transient displacement envelopes. A key challenge lies in the first-order differentiation of the transient amplitude with respect to time, as outlined in Eq.(5). Direct numerical differentiation of data points from Eq.(3) or similar peak-finding methods would introduce significant fluctuations in aerodynamic damping. Thus, selecting an appropriate trial function is essential. Fig.13 compares the impact of different trial functions:

- Higher-order polynomial functions for heaving VIV of a rectangular cylinder, with manually adjusted orders for optimal fitting, yield results comparable to Fig. 4 (which uses a second-order exponential decay function), though with slightly greater dispersion for $Sc$=78.1.
- Cubic spline functions produce notable fluctuations in the estimated Griffin plot lines.
- Higher-order polynomial functions for torsional VIV of a double-girder section, even after optimization, distort the estimated Griffin plots for Scruton numbers starting from 2~3, with a reduced effective range compared to Fig. 9.

Careful selection of the trial function is therefore necessary, and the second-order exponential decay function in Eq. (4) is recommended for its reliability.

(5) Reduced velocity considerations

Griffin plots do not correspond to a specific reduced velocity; peak VIV amplitudes for different Scruton numbers may occur at different reduced velocities. If the reduced velocity at peak amplitude changes significantly with Scruton numbers, the resulting Griffin plot may exhibit inaccuracies when extrapolated to distant Scruton numbers, as amplitude-dependent aerodynamic damping also varies with reduced velocity. Fortunately, for the cases in Fig.6a, 7a, and 9a, the variation in peak-amplitude reduced velocities with Scruton number is small. However, Fig.1 shows a clear variation, possibly because the reduced velocities were calculated using the still-air oscillation frequency $f_0$ rather than the frequency $f$ at specific wind velocities. Using the latter would likely mitigate this issue.

(6) Influence of structural nonlinearity in the spring-suspended system

Estimation amplitude-dependent aerodynamic damping using Eq.(6) requries consideration of the structural damping ratio. To ensure consistency across different structural damping conditions, the influence of structural nonlinearity in the spring-suspended oscillatory system and external dampers should be accounted for, particularly when the amplitude-dependency of the structural damping ratio is significant and varies with Scruton number. In such cases, the constant structural damping ratio in Eq.(6)

should be replaced with an amplitude-depedent one. For a comprehensive method to estimate amplitude-dependent structural damping, see Gao and Zhu (2015).

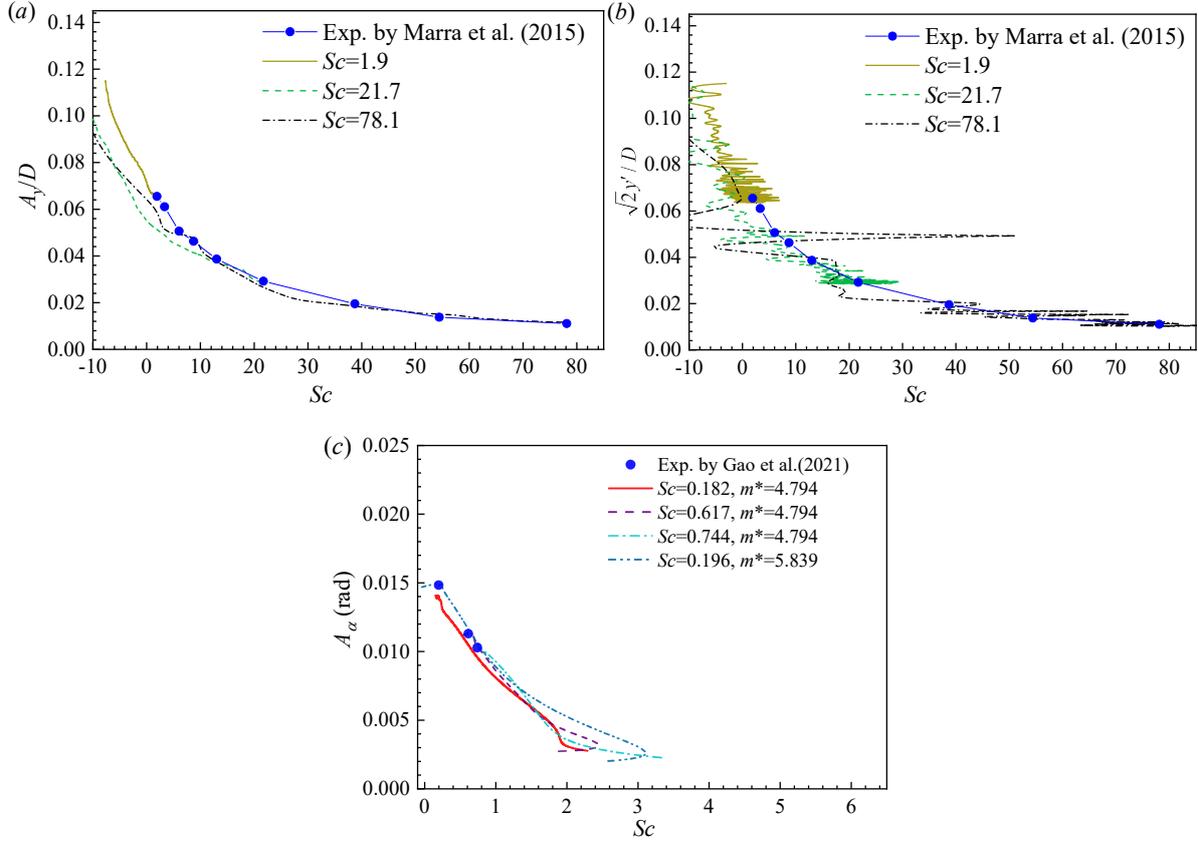

Fig. 13. Comparison of different trial functions in fitting the data of transient amplitude obtained by Eq.(3). (a) Higher-order polynomial function for the heaving VIV of a rectangular cylinder with a side ratio of 4:1, (b) Cubic spline function for the heaving VIV of a rectangular cylinder with a side ratio of 4:1, (c) Higher-order polynomial function for the torsional VIV of the double-girder section.

## 4. Empirical modelling of vortex-induced forces

As noted in Section 3.3, Griffin plots for typical bluff bodies demonstrate an approximately linear relationship between the reciprocal of stable VIV amplitudes and the Scruton number. This qualitative trend suggests that an empirical model of vortex-induced forces can be formulated to preserve this linearity in the theoretical expression of the Griffin plot. In this section, a simplified empirical VIV model is introduced, along with a method for identifying the associated parameters.

### 4.1. A novel VIV model to reproduce the Griffin plot

The linear relationship between the reciprocal of VIV amplitudes and the Scruton number mathematically resembles the influence of a Coulomb friction term on structural damping, as derived by Gao and Zhu (2015). To replicate this behavior, we incorporate a nonlinear Coulomb friction term into the model of nonlinear aerodynamic damping. For heaving VIV, the empirical model of the vortex-induced force is given by:

$$f_{\text{VIV}} = \rho U^2 D \left\{ Y_1(K) \left[ \frac{\dot{y}}{U} - \varepsilon(K) \cdot \text{sign}(\dot{y}) \right] + Y_2(K) \frac{y}{D} + \tilde{C}_L(K) \sin(\omega_{vs} t + \psi) \right\} \quad (9)$$

where $f_{\text{VIV}}$ represents aerodynamic force per unit length, $U$ is the mean oncoming velocity, $\dot{y}$ is heaving velocity, sign() is the sign function. $Y_1$, $\varepsilon$, $Y_2$ and $\tilde{C}_L$ are aerodynamic parameters to be estimated from wind tunnel tests, and they are dependent on reduced frequency $K = \omega D / U$ being $\omega$ the circular frequency under wind velocity $U$. $\omega_{vs}$ is the circular frequency of vortex shedding, and $\psi$ is the phase difference of vortex shedding.

Using the method of multiple scales (Gao et al., 2021), the transient amplitude resulting from an initial perturbation can be derived from Eq.(9) as:

$$\frac{\dot{a}}{a} = -K_0 \xi_s + \frac{m_r Y_1}{2} - \frac{2 m_r Y_1 \varepsilon}{\pi K} \frac{1}{a} \tag{10}$$

where $a = A_y / D$ is the dimensionless transient amplitude, $m_r = \rho D^2 / m$ represents dimensionless mass.

At steady state, when the transient amplitude converges to the stable amplitude, the time derivative $\dot{a}$ equals zero. Setting the right-hand side of Eq.(10) to zero yields the Griffin plot as

$$A_y^* = \frac{A_y}{D} = \frac{4\varepsilon}{\pi K} \cdot \frac{1}{1 - \frac{B}{D} \cdot \frac{K_0 Sc}{2\pi Y_1}} \tag{11}$$

An analogous empirical model for torsional VIV can be expressed as

$$M_{\text{VIV}} = \rho U^2 B^2 \left\{ Y_1(K) \left[ \frac{B\dot{\alpha}}{U} - \varepsilon(K) \cdot \text{sign}(\dot{\alpha}) \right] + Y_2(K)\alpha + \tilde{C}_M(K) \sin(\omega_{vs} t + \psi) \right\} \tag{12}$$

where $M_{\text{VIV}}$ represent the aerodynamic torsional moment per unit length, $\dot{\alpha}$ is torsional velocity. $\tilde{C}_M$ is the reduced frequency-dependent aerodynamic parameter of vortex shedding.

### 4.2. Stability analysis of resulting VIV amplitude from empirical modeling

The Griffin plot in Eq.(11) can be transformed into the following expressions for the stable VIV amplitude:

$$\frac{D}{A_y} = \frac{\pi K}{4\varepsilon} - \frac{K_0 K}{8 Y_1 \varepsilon} \frac{B}{D} Sc \tag{13}$$

It is evident from Eq.(13) that the proposed empirical model as Eq.(9) successfully captures the observed linear relationship between the reciprocal of VIV amplitude and Scruton number.

To examine the stability of the VIV amplitude prediced by the proposed model, we subsitute Eq.(10) into Eq.(5), yielding the total damping ratio as a function of transient amplitude:

$$\xi_{\text{total}}(a) = -\frac{1}{K} \frac{\dot{a}}{a} = \frac{K_0 \xi_s}{K} - \frac{m_r Y_1}{2K} + \frac{2 m_r Y_1 \varepsilon}{\pi K^2} \frac{1}{a} \tag{14}$$

The resultant VIV amplitude $A_y$ of Eq.(13) corresponds to the fixed point of Eq.(10), where the time derivative $\dot{a} = 0$, which is also the zero crossing point of the total damping $\xi_{\text{total}}(a)$. Inferring from the first equality in Eq.(14), when $\xi_{\text{total}} < 0$, the amplitude increases ($\dot{a} > 0$), indicating oscillation growth. Conversely, when $\xi_{\text{total}} > 0$, the amplitude decreases ($\dot{a} < 0$), signifying oscillation decay. Additionally, Fig.10~Fig.12 reveal a positive linear slope between reciprocal of amplitude $D/A_y$ and the Scruton number. For this relationship to hold in Eq.(13), the coefficient $Y_1 \varepsilon$ must be negative. As a result, a representative curve of total damping can be drawn in Fig.14, it is clear that the zero crossing point $A_y$ of the total damping $\xi_{\text{total}}(a)$ can be identified as a stable fixed point, consistent with experimental observations.

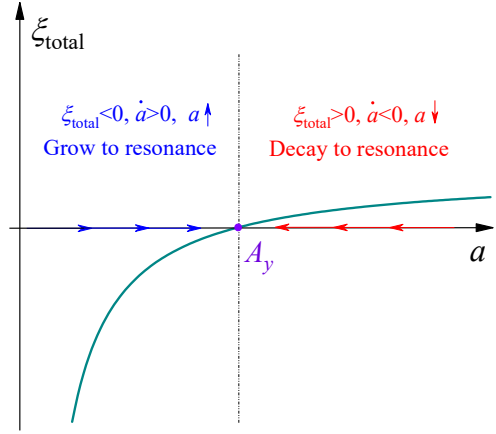

Fig. 14. Stability analysis of the fixed point derived from the amplitude-dependent total damping.

Griffin plots arise fundamentally from the aerodynamic nonlinearity. A purely linear empirical model would predict either zero amplitude or unbounded growth. The specific shape of the predicted Griffin plot depends on the mathematical form of the VIV empirical model. An interesting observation from Eq.(14) is that at small amplitude, the nonlinear term in Eq.(9) dominates the negative aerodynamic damping. Specifically, as amplitude approaches zero, the third term in Eq.(14) becomes infinitely large and provides a very large negative damping. This will result in a rapid transient amplitude, but this steep growth is tempered by reduced spanwise coherence of the vortex-induced forced at small amplitudes, limiting the growth rate. At large amplitudes, the linear aerodynamic damping term in Eq.(9) becomes dominant, as the third term in Eq.(14) tends to zero, leaving the second term to primarily govern the aerodynamic damping. This observation contrast with the the tranditional perspective, which suggests that linear aerodynamic damping term dominates the small-amplitude oscillation, while the nonlinear aerodynamic damping term become significant at large amplitude (Ehsan and Scanlan, 1990; Zhu et al., 2017). Here, the empirical model suggests the opposite: nonlinear effects drive small-amplitude dynamics, and linear effects stabilize large-amplitude responses. These contradictory views need further investigation.

### 4.3. Identification of aerodynamic parameters

The aerodynamic parameters $Y_1$ and $\varepsilon$ in Eq.(9) are associated with aerodynamic damping and can be determined from the estimated amplitude-dependent aerodynamic damping and stable amplitudes. Firstly, substituting Eq.(14) into the dimensionless form of Eq.(6), we obtain the expression for amplitude-dependent aerodynamic damping:

$$\xi_{\text{aero}}(a) = -\frac{m_r Y_1}{2K} + \frac{2m_r Y_1 \varepsilon}{\pi K^2}\frac{1}{a} \tag{15}$$

To determine the parameter $Y_1\varepsilon$, the estimated data points $\xi_{\text{aero}}(a_i)$, derived from transient displacement responses, are linearly fitted to Eq.(15). The fitted linear slope provides

$$Y_1\varepsilon = \frac{\pi K^2}{2m_r Y_1 \varepsilon}\frac{d\xi_{\text{aero}}}{d(a^{-1})} \tag{16}$$

Using Eq.(16), we can obtain $Y_1\varepsilon$ under any tested reduced velocity in the VIV lock-in range. At the peak amplitude state, $Y_1\varepsilon$ can be alternatively estimated by linearly fitting the curves of Griffin plot in Eq.(11) using the data points estimated by the method in Section 3.1, yielding

$$Y_1\varepsilon = -\frac{KK_0}{8}\frac{B}{D}\frac{\mathrm{d}\,Sc}{\mathrm{d}(D/A_y)} \tag{17}$$

With $Y_1\varepsilon$ known, we further obtain $Y_1$ from the stable amplitude $A_y$, by substituting $A_y$ into Eq.(10) and setting $\dot{a}=0$, yields

$$Y_1 = \frac{1}{2\pi}\left(\frac{8Y_1\varepsilon}{K}\frac{D}{A_y} + \frac{B}{D}K_0 Sc\right) \tag{18}$$

Parameter $\varepsilon$ is calculated by dividing the estimated $Y_1\varepsilon$ by $Y_1$. Under peak amplitude state, we have

$$\varepsilon = \frac{1}{\dfrac{K_0 Sc}{2\pi}\dfrac{B}{D}\dfrac{1}{Y_1\varepsilon} + \dfrac{D}{A_y}\dfrac{4}{\pi K}} \tag{19}$$

The aerodynamic coefficient $Y_2$ of the linear stiffness term in Eq.(9) can be inferred from the variation of oscillation frequency

$$Y_2 = \frac{K_0^2 - K^2}{m_r} \tag{20}$$

The remaining aerodynamic parameters $\tilde{C}_L$, $\omega_{vs}$ and $\psi$ are related with pure vortex shedding. As they typically have a negligible effect on the Griffin plots, they are thus not addressed in this study.

### 4.4. Comparison of calculated Griffin plots

Following the method outlined in Section 4.3, we determined the aerodynamic parameters of the proposed empirical model for the rectangular cylinder with a side ratio of 4:1. The results for three representative Scruton numbers are listed in Table 5. The identified values of $\varepsilon$ are consistent across different cases, while the values of $Y_1$ very, possibly due to the differences in the reduced frequency at peak amplitude states across the Scruton numbers.

Substituting the identified aerodynamic parameters into Eq.(11), we can generate the Griffin plots using our empirical model. The results are illustrated in Fig.15a and compared with experimental data by Marra et al. (2015). The generated curves of Griffin plots under three representative Scruton numbers are in satisfactory agreement with experimental results, although minor dispcrepancies are observed at low Scruton numbers. This suggests that the proposed VIV model can accurately capture the general trend of Griffin plots when aerodynamic parameters are estimated from the transient displacement response at a single Scruton number.

For comparison, we also generate curves of Griffin plots using the Van der Pol type model proposed by Ehsan and Scanlan (1990), for which the Griffin plot is expressed as (Marra et al., 2015):

$$\frac{A_y}{D} = \frac{2}{\sqrt{\varepsilon}}\sqrt{1 - \frac{B}{D}\cdot\frac{K_0 Sc}{2\pi Y_1}} \tag{21}$$

where parameter $\varepsilon$ is related a nonlinear damping term characteristic of the Van der Pol type model.

The results are presented in Fig.15b, we observed that Ehsan and Scanlan's model fails to preserve the shape of Griffin plots when the Scruton number deviates from the one for parameter identification. Ehsan and Scanlan (1990) recognized this limitation as their identified aerodynamic parameters show a strong dependency on the structural damping parameters. Recently, Marra et al. (2011, 2015) addressed this issue by extending the aerodynamic parameters in Ehsan and Scanlan's model to depend on the Scruton number, though this approach requires more testing to calibrate the dependency.

Table 5. Identified aerodynamic parameters of the proposed empirical model. $\mathrm{d}(D/A_y)/\mathrm{d}Sc$ means the linear slope in Fig.10a. $A_y/D$ represents the dimensionless stable VIV amplitude.

| $Sc = \dfrac{4\pi m \xi_s}{\rho BD}$ | $m_r = \dfrac{\rho D^2}{m}$ | $\mathrm{d}(D/A_y)/\mathrm{d}Sc$ | $K = \omega D/U$ | $A_y/D$ | $Y_1$ | $\varepsilon$ |
|---|---|---|---|---|---|---|
| 1.9 | 9.442×10⁻⁴ | 0.8927 | 0.6549 | 0.06551 | -6.337 | 0.03791 |
| 21.7 | 9.411×10⁻⁴ | 0.7543 | 0.7364 | 0.0292 | -11.111 | 0.03235 |
| 78.1 | 9.411×10⁻⁴ | 0.9583 | 0.7733 | 0.01106 | -8.001 | 0.03900 |

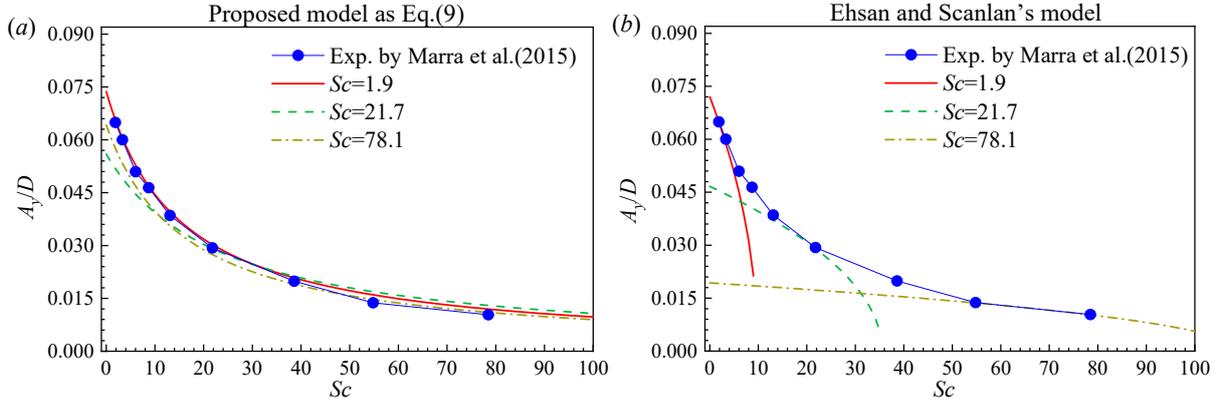

Fig. 15. Comparison of predicted Griffin plot and experimental data where lines represent the aerodynamic parameters are identified from the transient response under the corresponding Scruton number. (a) Proposed model, (b) Ehsan and Scanlan's model.

Apart from the Van der Pol-type model proposed by Ehsan and Scanlan, previously developed VIV empirical models incorporating nonlinear Rayleigh-type damping terms or hybrid Van der Pol and Rayleigh types also fail to adequately preserve the general trend of Griffin plots. Because the Rayleigh type models are equivalent to the Van der Pol type models in terms of amplitude-dependent aerodynamic damping (Gao et al., 2021). The applicability of other existing models warrants further investigation. Notably, most empirical models lack rigorous validation of their ability to reproduce Griffin plots across a broad range of Scruton numbers, particularly at values significantly distant from the conditions used for parameter identification. Indeed, the capacity to accurately reconstruct Griffin plots with minimal experimental data is a fundamental requirement for any practical empirical model. Despite its simplicity, our proposed empirical model demonstrates a notable ability to extrapolate to Scruton numbers far beyond the tested parameters.

## 6. Conclusions

This paper invesigates the self-similarity of Griffin plots for vortex-induced vibration (VIV) in air flow. Based on this self-similarity, a numerical method is proposed to estimate Griffin plots from transient displacement responses. The method is applied to experimental data of three typical bluff sections to construct Griffin plots and validate the feasibility of self-similarity. Then, a novel VIV empirical model is developed, along with a method for identifying its parameters. The original contributions and major conclusions are summarized as follows:

- We demonstrate that Griffin plots exhibit self-similarity, reflecting a strong connection to amplitude-dependent aerodynamic damping. This property enables estimation of Griffin plots from transient displacement responses at any Scruton number.
- The proposed numerical method facilitates rapid estimation of Griffin plots from transient displacements. The resulting plots, derived from various Scruton numbers, align closely with each other and match experimental values.
- Self-similarity in Griffin plots is confirmed for both cross-flow and torsional VIV across typical bluff sections. Apart from cross-flow VIV, Scruton number also serves as a unified mass-damping parameter for torsional VIV.
- For rectangular sections and typical flat bridge decks, Griffin plots display a consistent pattern: the reciprocal of VIV amplitudes shows an approximately linear relationship with the Scruton number, particularly in torsional VIV.
- The proposed empirical model of vortex-induced forces accurately reproduces the Griffin plot using aeroelastic parameters identified from experimental data at a single Scruton number, significantly reducing the need for extensive experimental measurements.

Finally, this study establishes a foundation for future VIV modeling in marine engineering. However, the conclusions presented here require further validation using experimental data from water tunnel tests.

## Acknowledgments

The funding was provided by National Natural Science Foundation of China, 52278478, Research Funds for the Interdisciplinary Projects, CHU (grant No. 300104240923), by the Fundamental Research Funds for the Central Universities, CHD (Grand No. 300102214914). The authors also wish to thank Antonino Marra, Claudio Mannini for the sharing the experimental data of the rectangular cylinder.

# References

Chen, C., Mannini, C., Bartoli, G., Thiele, K., 2022. Wake oscillator modeling the combined instability of vortex induced vibration and galloping for a 2:1 rectangular cylinder. J. Fluids Struct. 110, 103530.

Ehsan, F., & Scanlan, R.H., 1990. Vortex-induced vibrations of flexible bridges. J. Eng. Mech. 116, 1392–411.

Farshidianfa, A., & Zanganeh, H., 2010. A modified wake oscillator model for vortex-induced vibration of circular cylinders for a wide range of mass-damping ratio. J. Fluids Struct. 26, 430-441.

Francisco, H.H., 2025. Vortex-induced vibration of flexible cylinders in cross-flow. Annu. Rev. Fluid Mech. 57, 285-310.

Gao, G.Z., Zhu, L.D., 2015. Nonlinearity of mechanical damping and stiffness of a spring-suspended sectional model system for wind tunnel tests. J. Sound Vib. 355, 369–391.

Gao, G.Z., Zhu, L.D., Li, J.W., Han, W.S., 2020. Application of a new empirical model of nonlinear self-excited force to torsional vortex-induced vibration and nonlinear flutter of bluff bridge sections. J. Wind Eng. Ind. Aerodyn. 205, 104313.

Gao, G.Z., Zhu, L.D., Bai, H., Han, W.S., Wang, F., 2021. Analytical and experimental study on Van der Pol-type and Rayleigh-type equations for modeling nonlinear aeroelastic instabilities. Adv. Struct. Eng. 24, 1-20. https://doi.org/10.1177/13694332211022056

Ge, Y.J., Zhao, L., Cao, J.X., 2022. Case study of vortex-induced vibration and mitigation mechanisms for a long-span suspension bridge. J. Wind Eng. Ind. Aerodyn. 220, 104866.

Govardhan, R., & Williamson, C.H.K., 2006. Defining the 'modified Griffin plot' in vortex-induced vibration: revealing the effect of Reynolds number using controlled damping. J. Fluid Mech. 561, 147-180.

Hansen, S.O., 2013. Vortex-induced vibrations – the Scruton number revisited. Structures and Buildings 166, 560-571. http://dx.doi.org/10.1680/stbu.11.00018

Hartlen, R.T., Currie, I.G., 1970. Lift-oscillator model of vortex-induced vibration. J. Eng.Mech. Division (ASCE), 96 (5), 577–591.

Hui, Y., Tang, Y., Yang, Q., Chen, B., 2024. A wake-oscillator model for predicting VIV of 4-to-1 rectangular section cylinder. Nonlin. Dyn., 112, 8985–8999. https://doi.org/10.1007/s11071-024-09516-9

Hwang, Y.C., Kim, S., Kim, H.K., 2020. Cause investigation of high-mode vortex-induced vibration in a long-span suspension bridge. Structure and Infrastructure Engineering, 16:1, 84-93. DOI: 10.1080/15732479.2019.1604771

Iwan, W., Botelho, D., 1985. Vortex-induced oscillation of structures in water. Journal of Waterway, Port, Coastal, and Ocean Engineering ASCE 111 (2), 289–303.

Khalak, A., Williamson, H., 1999. Motions, forces and mode transitions in vortex-induced vibrations at low mass–damping. J. Fluids Struct. 13(7–8), 813–851.

Kim, H.K., Hwang, Y.C., Kim, S.J., 2016. Unexpected vibration monitoring in a suspension bridge and cause investigation. In: Proceedings of the 2016 Structures Congress (Structures16), Jeju Island, Korea.

Larsen, A., 1995. A generalized model for assessment of vortex-induced vibrations of flexible structures. J. Wind Eng. Ind. Aerodyn. 57 (2–3), 281–294.

Li, J.W., Yang, S.C., Hao, J.M., Gao, G.Z., Wang, F., Bai, H., Zhao, G.H., Li, Y., Xue, X.F., 2024. Advances and applications of wind engineering in exceptional terrain. J. Traffic Transp. Eng. (Engl. Ed.). 11(6), 1023-1209. https://doi.org/10.1016/j.jtte.2024.09.002

Liao, H.L. https://weibo.com/ttarticle/p/show?id=2309404681020849914139, 2021.

Marra, A.M., Mannini, C., Bartoli, G., 2011. Van der Pol-type equation for modeling vortex-induced oscillations of bridge decks. J. Wind Eng. Ind. Aerodyn. 99, 776-785.

Marra, A.M., Mannini, C., Bartoli, G., 2015. Measurements and improved model of vortex-induced vibration for an elongated rectangular cylinder. J. Wind Eng. Ind. Aerodyn. 147, 358-367.

Qie, K., & Zhang, Z.T., 2024. Nonlinear effects of mechanical and aerodynamic damping on a motion-amplitude-dependent model. International Journal of Structural Stability and Dynamics, https://doi.org/10.1142/S0219455425501202

Qu, C.X., Tu, G.K., Cao, F.Z., Sun, Li., Pan, S.S., Chen, D.S., 2024. Review of bridge structure damping model and identification method. Sustainability, 16, 9410. https://doi.org/10.3390/su16219410

Sarpkaya, T., 1978. Fluid forces on oscillating cylinders. Journal of Waterway, Port, Coast, and Ocean Division, ASCE 104, 275–290.

Sarpkaya, T., 1979. Vortex-induced oscillations. ASME J. Appl. Mech. 46, 241–258.

Scruton, C., 1963. On the wind-excited oscillations of stacks, towers and masts. In: Proceedings of the International Conference on the Wind Effects on Buildings and Structures, Teddington, Middlesex, 1963, pp. 798–837.

Scruton, C., 1965. On the wind-excited oscillations of towers, stacks and masts. In: Proceedings of the Symposium Wind Effects on Buildings and Structures, vol.16, HMSO, London, pp. 798–836.

Skop, R., Griffin, O., 1973. A model for the vortex-excited resonant response of bluff cylinders. J. Sound Vib. 27 (2), 225–233.

Skop, R.A., 1974. On modelling vortex-excited oscillations. NRL Memo. Rep. 2927.

Tan, J.F., Chu, X.L., Cui, W., Zhao, L., 2024. Life-cycle assessment for flutter probability of a long-span suspension bridge based on operational monitoring data. Journal of Infrastructure Intelligence and Resilience, 3, 100108.

Vandiver, J.K. 2002. A universal reduced damping parameter for prediction of vortex-induced vibration. In: 21st International Conference on OMAE, June 23-28, Oslo.

Vickery, B.J., Watkins, R.D., 1964. Flow-induced vibrations of cylindrical structures. In: Silvester, R. (Ed.), Proceedings of the First Australian Conference on Hydraulics and Fluid Mechanics. Pergamon, New York, pp. 213–241.

Williamson, C.H.K., & Govardhan, R., 2004. Vortex-induced vibrations. Annu. Rev. Fluid Mech. 36, 413-55. Doi: 10.1146/annurev.fluid.36.050802.122128.

Williamson, C.H.K., & Govardhan, R., 2008. A brief review of recent results in vortex-induced vibrations. J. Wind Eng. Ind. Aerodyn. 96, 713-735.

Wu, B.C., & Laima, S.J., 2021. Experimental study on characteristics of vortex-induced vibration of a twin-box girder and damping effects. J. Fluids Struct. 103, 103282.

Zasso, A., Belloli, M., Giappino, S., Muggiasca, S., 2008. Pressure field analysis on oscillating circular cylinder. J. Fluids Struct. 24, 628-650.

Zhang, M.J., Petersen, Ø. W., Øiseth, O.A., 2024. Identification of amplitude-dependent aerodynamic damping from free vibration data using iterative unscented kalman filter. J. Wind Eng. Ind. Aerodyn. 253, 105850.

Zdravkovich, M.M., 1982. Modification of vortex shedding in the synchronization range. ASME J. Fluids Eng. 104, 513–517.

Zdravkovich, M., 1990. On origins of hysteretic responses of a circular cylinder induced by vortex shedding. Flugwissenschaften Weltraumforschung 14, 47–58.

Zhu, L.D., Meng, X.L., Du, L.Q., Ding, M.C., 2017. A simplified nonlinear model of vertical vortex-induced force on box decks for predicting stable amplitudes of vortex-induced vibrations. Engineering 3, 854-862.

**Nomenclature**

- $a$ Dimensionless transient amplitude $a = A_y / D$
- $A$ Transient displacement
- $\bar{A}, \dot{\bar{A}}$ Transient amplitude and its time derivative with high-frequency noise filtered out
- $A_y$ The steady amplitude of cross-flow VIV
- $A_\alpha$ The steady amplitude of torsional VIV
- $B$ Section width
- $\tilde{C}_L, \tilde{C}_M, \psi$ Amplitude and phase angle of the vortex shedding term
- $D$ Section depth
- $f$ Frequency in flowing air condition
- $f_0$ Natural frequency
- $J_m$ Effective mass moment of inertia per unit length
- $K, K_0$ Reduced frequency $K = \omega D / U, K_0 = \omega_0 D / U$
- $m$ Effective mass per unit length
- $m_r$ Mass ratio $m_r = \rho D^2 / m$
- $m^*$ Dimensionless mass or dimensionless mass moment of inertia
- $N$ Total number of sampled points
- $s$ Dimensionless time $s = Ut / D$
- $Sc$ Scruton number
- $t$ Time
- $U$ Mean velocity in free stream
- $U^*$ Reduced velocity $U^* = U / (fD)$
- $y, \dot{y}$ Heaving displacement, velocity
- $y'$ Root mean square (RMS) of the steady-state displacement
- $Y_1, Y_2$ Linear aerodynamic parameters
- $\Delta t$ The sampling time interval.
- $\varepsilon$ Aerodynamic paramerer related with nonlinear damping term
- $\theta_0$ Phase angle of displacement
- $\xi_{aero}$ Aerodynamic damping ratio
- $\xi_s$ Structural damping ratio
- $\xi_{total}$ Total damping ratio
- $\rho$ Air density
- $\omega$ Circular frequency